# Towards a Security-Aware Benchmarking Framework for Function-as-a-Service

Roland Pellegrini[1], Igor Ivkic[1] and Markus Tauber[1]

[1]*University of Applied Sciences Burgenland, Eisenstadt, Austria*
*{1610781022, igor.ivkic, markus.tauber}@fh-burgenland.at*

Abstract: In a world, where complexity increases on a daily basis the Function-as-a-Service (FaaS) cloud model seams to take countermeasures. In comparison to other cloud models, the fast evolving FaaS increasingly abstracts the underlying infrastructure and refocuses on the application logic. This trend brings huge benefits in application and performance but comes with difficulties for benchmarking cloud applications. In this position paper, we present an initial investigation of benchmarking FaaS in close to reality production systems. Furthermore, we outline the architectural design including the necessary benchmarking metrics. We also discuss the possibility of using the proposed framework for identifying security vulnerabilities.

## 1 INTRODUCTION

Cloud computing, as defined by Mell and Grance (2011), is a model for enabling on-demand network access to a shared pool of configurable resources. Cloud vendors provide these resources in the service models Infrastructure as a Service (IaaS), Platform-as-a-Service (PaaS) and Software-as-a-Service (SaaS). Through virtualization, the IaaS service model provides computing resources (e.g.: servers, storage, network) to consumers so they can deploy and run software. In other words, the consumers do not control the infrastructure, but are able to manage the running operating systems and applications. Quite contrary to IaaS, PaaS offers an integrated runtime and development environment where consumers only control their deployed and developed applications. Finally, SaaS provides software and applications, which can be used and accessed by consumers via the web or application programming interfaces (API).

These three service models are currently being extended by a very new and rapidly evolving technology called Function-as-a-Service (FaaS). FaaS provides a runtime environment to develop, deploy, run, and manage application functionality without any knowledge about the underlying application. All instances of these environments are managed by the provider, who is responsible for the code execution, resource provisioning and automatic scaling for virtually any type of application. Figure 1 gives an overview of a generic FaaS architecture and illustrates the technical workflow between the components:

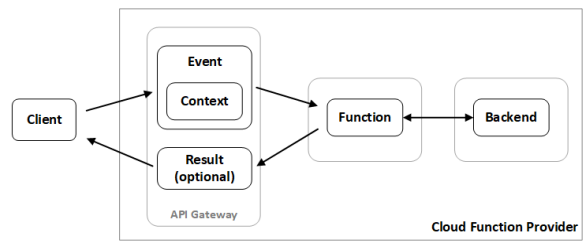

Figure 1: FaaS architecture, based on Pientka (2017)

However, Cloud Service Providers (CSP) often limit the amount of execution time or resource allocation a request may consume. Additionally, FaaS code may suffer more from start-up latency than code that is continuously running on a dedicated server. The reason for that is that each time a function is called the underlying environment has to be provisioned. Depending on the providers configuration, the function may wait an amount of time before it is deprovisioned. If another request is sent to the function while it is waiting it executes the request again. But, in case the function has already been deprovisioned and is reinvoked by an event, the runtime environment has to start up again, which leads to delays and latency.

As explained in Section II in more detail, the existing publications either compare CSPs and/or technologies, or use approaches, which are applicable for specific use cases. In order to make FaaS benchmarking possible in close to reality production systems and independent of a specific use case a more general approach is required.

In this position paper we propose a framework for benchmarking FaaS in production systems, which is independent of a specific use case. In this regard we present the archictectual design, discuss two different methods of measuring FaaS performance, propose a set of benchmarking metrics and explain, which components of the benchmarking framework need to be implemented.

The rest of the position paper is structured as follows: Section II provides a summary of the related work, followed by the benchmark framework archictechture in Section III. Finally, we present our conclusions and future work including security considerations in Section IV.

## 2 RELATED WORK

In the field of cloud performance evaluation most of the existing research publications can be grouped in the following three categories: (a) IaaS, PaaS, and SaaS benchmarking at different abstraction levels, (b) comparison of existing Cloud Benchmark Frameworks, and (c) performance comparison of cloud services among CSPs.

Since FaaS is a relatively new technology, more detailed research is needed for benchmarking cloud functions. An initial introduction and guideline has been done by Bermbach et al. (2017) where they compare the cloud service models, define terms related to benchmarking, derive the quality requirements of the users, and explain the motivation for benchmarking these qualities. Additionally, they cover the entire lifecycle of cloud service benchmarking, from its motivations, over benchmarking design and execution, to the use of benchmarking results.

Sitaram and Manjunath (2011) provide another introduction in cloud computing and examine some popular cloud benchmarking tools. Whereas, Mueller et al. (2014), Coarfa et al. (2006), and Juric et al. (2006) evaluated security performance for different use cases in experimental studies.

Malawski et al. (2017) focus on performance evaluation of cloud functions by taking heterogeneity aspects into account. For this purpose, they developed a framework with two types of CPU-intensive benchmarks for performance evaluation of cloud functions. Then they applied it to all the major Cloud Function Providers (CFP) such as Amazon, Microsoft, Google, and IBM. Their results show the heterogeneity of CFPs, the relation between function size and performance and how CFPs interpret the resource allocation policies differently.

Hwang et al. (2016) provided a summary of useful Cloud Performance Metrics (CPM) and introduced an extended CPM-concept on three levels: (a) basic performance metrics including traditional metrics (e.g. execution time, speed and efficiency), (b) cloud capabilities for describing network latency, bandwidth, and data throughput, and finally, (c) cloud productivity, which deals with productivity metrics (e.g. Quality of Service (QoS), Service Level Agreement (SLA) and security).

Additionally, Luo et al. (2012) proposes a benchmark suite for evaluating cloud systems running data processing applications. Furthermore, they discussed and analysed, why traditional metrics (e.g. floating-point operations and I/O operations) are not appropriate for system cloud benchmarking. Instead, they propose data processed per second and data processed per Joule as two complementary metrics for evaluating cloud computing systems.

The architecture of cloud benchmarking tools has been subject of previous research. An overview of a generic architecture, elements of a benchmarking tool and performance metrics for IaaS cloud benchmarking has been discussed by Iosup et al. (2014). Furthermore, Sangroya and Bouchenaket (2015) proposed a generic software architecture for dependability and performance benchmarking for cloud computing services. They also describe various components and modules responsible for injecting faults in cloud services in addition to the components responsible for measuring the performance and dependability.

Finally, the Yahoo! Cloud Serving Benchmark (YCSB) is a benchmark suite that measures the performance of a cloud storage system against standard workloads. The architecture, as described by Cooper et al. (2010), consists of a YCSB client, which is a Java-based multi-threaded workload generator and can be extended to support benchmarking different databases.

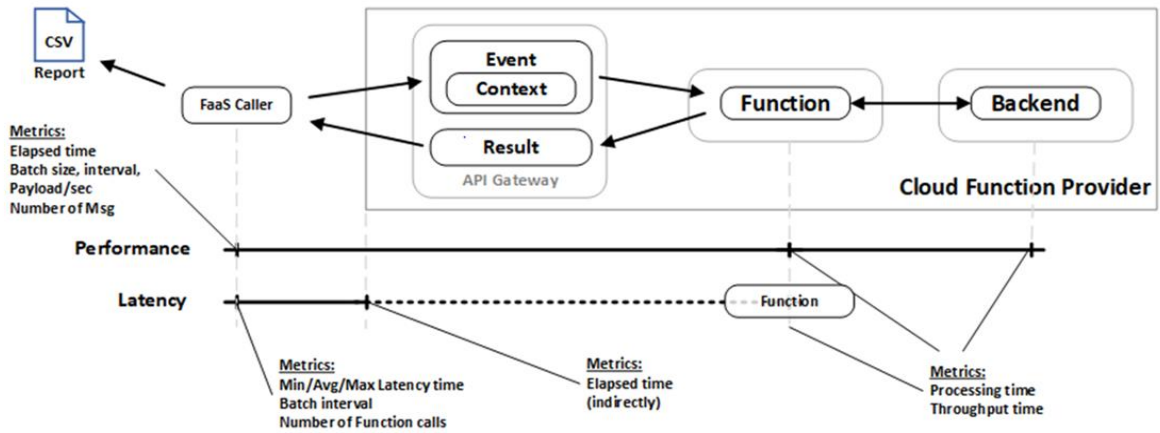

Figure 2: Prototype architecture, based on Pientka (2017)

# 3 PROPOSED ARCHITECTURE FRAMEWORK

The approach taken in this position paper differs from the existing publications as mentioned above. The main focus is to evaluate the FaaS performance for data processing (e.g. the maximum number of function calls per seconds) instead of identifying the hardware specification of the underlying IT infrastructure of the CFP. In order to evaluate FaaS performance the following influencing parameters need to be considered: (a) FaaS caller with varying parameters such as data size, run length, batch size, and pause interval, (b) cloud function latency and throughput, and (c) network latency and bandwidth.

In this position paper we propose an approach for benchmarking FaaS performance on a close to reality system without the need of implementing a complex testbed. To achieve this goal, the architecture of the benchmarking framework needs to be designed in a way, so that the production system needs a minimum amount of adaptation. The reason for that is that a benchmarking executed on a production system delivers more significant results in comparison to a test environment. In more detail, the benchmarking framework has to be designed as a two-tier architecture. In this way the FaaS calls will be executed on a sender and the cloud function processing will run on the CSP platform. We consider two FaaS sender variants: A Faas Performance Caller (FaaS-PC) and a FaaS Latency Caller (FaaS-LC).

The FaaS-PC measures the potential performance of a cloud function by sending runs of messages, and compiling statistics. Instead of sending a continuous stream of messages, the FaaS-PC groups them into batches. Between two batches the FaaS-PC pauses for a freely configurable interval of seconds. Furthermore, the FaaS-PC supports a single mode and an automatic mode. In the single mode, the FaaS-PC sends a single run of messages. This mode is useful to answer questions about cloud function behavior under sustained load conditions. In the automatic mode the FaaS-PC starts sending several runs of messages and function calls, modifying the parameters for each run. It does so, until the ideal batch size and interval parameters that yield maximum sustainable cloud function throughput are found. This mode allows FaaS-PC to tune its send rate to match the maximum receive rate of the slowest cloud function call-back. Both, the single mode and the automatic mode provide a summary-report at the end of the run.

In contrast to FaaS-PC, the FaaS-LC helps to identify latency bottlenecks in networks where the transit time between FaaS caller and cloud function needs to be kept to a minimum. However, clock synchronization between FaaS-LC and cloud function is not precise enough to accurately measure one-way travel time. Therefore, FaaS-LC measures round-trip time for a request-reply message pair by using a single clock. For this scenario an additional function needs to be developed.

The following metrics are potential candidates to measure the performance for both types of FaaS callers: elapsed time, number of cloud function calls, number of messages, size of payloads, total payload size, batch interval, batch size, messages per second, payload per second (in bytes), maximum latency (in milliseconds), minimum latency (in milliseconds), number of messages or function calls with latency exceeds a threshold.

In the end, both FaaS callers generate a report, which can easily be transferred to spreadsheet applications or command-line programs. Another benefit is that two- and three-dimensional plots of functions, data and statistical reports can be produced. In Figure 2 the prototype architecture for FaaS benchmarking, the planned measuring-points, and metrics candidates are shown. As shown in the dotted line, the elapsed time has to be measured indirectly between the gateway and the function if the access to the API gateway is not possible.

## 4 CONCLUSION AND FUTURE WORK

In this position paper, we introduced a FaaS benchmarking framework for measuring the cloud function performance in a production environment for a front-to-back processing. First, we compared the functionality of FaaS to the traditional cloud service models. Next, we explained the technical architecture of FaaS and pointed out some related performance issues. In this regard, we discussed several aspects of cloud benchmarking and cloud security. Finally, in Section III we proposed a prototype including the architectural design and the functional requirements. In this regard, we outlined the necessity of the FaaS-PC and FaaS-LC from the benchmarking perspective. In addition to that we identified a set of metrics for measuring the performance of these FaaS callers.

In summary, we explored the possibility of measuring the performance of FaaS to make CSPs more comparable. By doing so, we provided a method for decision-makers, IT architects and cloud service consumers to assist them in finding the best FaaS solution for their businesses. The main contribution of this paper is the initial investigation on an approach for benchmarking FaaS, which can also be used to identify FaaS security vulnerabilities. While FaaS dramatically reduces some top-level threats, there are still some risks regarding DoS attacks and exploitation of the long-lived FaaS container functionality. Even though the proposed framework is used to measure the performance, it could also be applied to stress indirectly the underlying IaaS, PaaS and/or SaaS to emulate e. g. DoS attacks. This would make the Cloud return error codes, which could be exploited as security vulnerabilities. In future work, we will consider using the proposed benchmarking framework to identify possible FaaS vulnerabilities, threats and attacks to verify a broader application of our work.

## ACKNOWLEDGEMENTS

The research has been carried out in the context of the project MIT 4.0 (FE02), funded by IWB-EFRE 2014 - 2010.